\documentclass[aip,amsmath,amssymb,reprint]{revtex4-1}

\usepackage{graphicx}
\usepackage{dcolumn}
\usepackage{bm}
\usepackage[utf8]{inputenc}
\usepackage[T1]{fontenc}
\usepackage{mathptmx}
\usepackage{xcolor}
\usepackage{longtable}
\usepackage{booktabs}
\begin{document}

\title{Study of self-interaction-errors in barrier heights using locally scaled and Perdew-Zunger self-interaction methods}
\author{Prakash Mishra}
\affiliation{Computational Science Program, University of Texas at El Paso, El Paso, Texas 79968}
\author{Yoh Yamamoto}
\affiliation{Department of Physics, University of Texas at El Paso, Texas, 79968}
\author{J. Karl Johnson}
\affiliation{Department of Chemical \& Petroleum Engineering,University of Pittsburgh,Pittsburgh,Pennsylvania 15261}
\author{Koblar A. Jackson}
\affiliation{Physics Department and Science of Advanced Materials Program, Central Michigan University, Mt. Pleasant, Michigan 48859}
\author{Rajendra R. Zope}
\author{Tunna Baruah}
\email{tbaruah@utep.edu}
\affiliation{Computational Science Program, University of Texas at El Paso, El Paso, Texas 79968}
\affiliation{Department of Physics, University of Texas at El Paso,Texas, 79968}
\date{\today}

\begin{abstract}
  We study the effect of self-interaction errors on the barrier heights of chemical reactions. For this purpose we use the well-known Perdew-Zunger [J. P. Perdew and A. Zunger, Phys. Rev. B, {\bf 23}, 5048 (1981)] self-interaction-correction (PZSIC), as well as two variations of the recently developed, locally scaled self-interaction correction (LSIC) [R. R. Zope \textit{et al.}, J. Chem. Phys. {\bf 151}, 214108 (2019)] to study the barrier heights of the BH76 benchmark dataset. Our results show that both PZSIC and especially the LSIC methods improve the barrier heights relative to the local density approximation (LDA).  The version of LSIC that uses the iso-orbital indicator $z$ as a scaling factor gives a more consistent improvement than an alternative version that uses an orbital-dependent factor $w$ based on the ratio of orbital densities to the total electron density.  We show that LDA energies evaluated using the self-consistent and self-interaction-free PZSIC densities can be used to assess density-driven errors. The LDA reaction barrier errors for the BH76 set are found to contain significant density-driven errors for all types of reactions contained in the set, but the corrections due to adding SIC to the functional are much larger than those stemming from the density for the hydrogen transfer reactions and of roughly equal size for the non-hydrogen transfer reactions.  
\end{abstract}
\maketitle

\section{\label{sec:level1}INTRODUCTION}
Density functional theory (DFT) calculations with approximate semi-local exchange-correlation functionals fail to predict certain properties such as band gaps, reaction barriers, and fragment dissociation energies\cite{perdew1985density,PhysRevLett.51.1888,patchkovskii2002improving,grafenstein2004impact,ruzsinszky2006spurious} due to the self-interaction error (SIE). It is known that SIE arises from the incomplete cancellation of the self-Coulomb interaction by the density functional approximation (DFA) of the self-exchange energy for a one-electron density. 
The effect of SIE is particularly noted for systems with stretched bonds such as transition states in chemical reactions. \cite{zhang1998challenge,patchkovskii2002improving,grafenstein2004impact,shahi2019stretched} With DFAs, the total energy of an $N$-electron system  deviates from piece-wise linearity between integer numbers of electrons. The total energy of an $N$-electron system varies as a convex curve as a function of charge between $N$ and $N+1$ electrons\cite{PhysRevLett.100.146401,cohen2008insights} which is known as the charge delocalization problem. Charge delocalization results in lowering the energies of the transition states resulting in underestimation of  reaction barriers.\cite{chermette2001posteriori} 
In this work we investigate the effect of one-electron self-interaction corrections on reaction barrier heights. 
 
The self-interaction correction method of Perdew and Zunger \cite{perdew1981self,zunger1980self} (PZSIC) employs an orbital-by-orbital correction scheme to remove the one electron SIE. The PZSIC total energy is given by
\begin{equation}\label{eq:pzsic}
    E^{PZSIC}[\rho_\uparrow,\rho_\downarrow] = E^{DFA}[\rho_\uparrow,\rho_\downarrow] - \sum_{i\sigma}^{occ}\{ U[\rho_{i\sigma}] + E_{XC}^{DFA}[\rho_{i\sigma},0] \}
\end{equation}
where $i$ is the orbital index, $\sigma$ is the spin index,  $\rho_{i\sigma}$ is the density of the $i^{th}$ orbital, and $U[\rho_{i\sigma}]$ and $E_{XC}[\rho_{i\sigma},0]$ are the self-Coulomb and self-exchange-correlation energies of the $i^{th}$ orbital. $E^{PZSIC}$ is orbital dependent and the total energy depends not only on the total density, but also on the specific choice of orbitals used to represent that density. Pederson \textit{et al.} \cite{pederson1984local} showed how to determine the localized orbitals that minimize $E^{PZSIC}$ and maintain size-extensivity in the theory. 
 
PZSIC has been adopted by several different groups in the past with both real\cite{szotek1991self,rieger1995self,PhysRevA.55.1765,PhysRevA.50.2191,grafenstein2004impact,polo2002electron, zope1999atomic,temmerman1999implementation,patchkovskii2001curing,garza2000optimized,luders2005self,PhysRevB.75.045101,daene2009self,PhysRevB.28.5992,vydrov2004,korzdorfer2008self,messud2008improved,vieira2010investigation, poilvert2015koopmans} and complex\cite{klupfel2012effect,doi:10.1021/acs.jctc.6b00347} local orbitals.
Although PZSIC performs well in describing properties that are strongly impacted by SIE,\cite{vydrov2004,sharkas2018shrinking,C9CP06106A,akter2020study} it tends to over-correct  equilibrium properties that are already described well by semi-local DFAs. This is known as the ``paradox of SIC.'' \cite{perdew2015paradox} The origin of the paradox has been linked to the effect of SIC on slowly-varying densities. \cite{santra2019perdew} Non-empirical local or semi-local density functionals such as the Perdew-Wang local density approximation (LDA) \cite{PhysRevB.45.13244} or the Perdew, Burke, and Ernzerhof (PBE) generalized gradient approximation (GGA)\cite{PhysRevLett.77.3865} are designed to be exact in the limit of uniform electron density, and it has been argued that satisfying this constraint is important for achieving accurate descriptions of molecules.\cite{kaplan2020simple} But when PZSIC is applied, the corrected DFAs violate this constraint.\cite{santra2019perdew}
Recently, Zope \textit{et al.} presented an alternative approach to correcting SIEs. \cite{doi:10.1063/1.5129533} In this approach, called the locally scaled SIC (LSIC) method, an iso-orbital indicator is used to determine the nature of the charge density at a given point in space, distinguishing one-electron-like regions from regions where the density is slowly-varying and many-electron-like. The self-exchange and self-Coulomb energy densities of each orbital are then scaled locally such that full SIC is maintained in one-electron regions, but scaled down in slowly-varying regions. LSIC mitigates the over-correcting tendency of PZSIC, resulting in accurate results for properties such as atomization energies, electron affinities, magnetic properties, dipole moments and polarizabilities.\cite{doi:10.1063/1.5129533,C9CP06106A,Kushanta_dipole_2021,akter2020study,akter2021static}
The PZSIC method  not only removes one electron SIE but it also nearly satisfies 
the Perdew–Parr–Levy–Balduz (PPLB) condition.\cite{doi:10.1063/1.2566637,perdew1982density} 
The PPLB  condition
demands that the total energy $E$ of a system as a function of electron number $N$ to be
piece-wise straight lines interpolating between the nearest integers. Most DFAs 
including hybrid functionals fail to satisfy the PPLB condition and 
the error introduced is known as many-electron self-interaction error 
or delocalization error.
  LSIC is, by definition, one electron SIE free  and is likely 
to be nearly many-electron self-interaction-free as it reduces to PZSIC in the one-electron 
and dissociation limit, and it provides a good description of polarizabilities and other properties
that depend on the asymptotic 
behavior of the potential.\cite{akter2020study,akter2021static,akter2021well,C9CP06106A}

In this work we examine the performance of PZSIC and LSIC for chemical reaction barriers.  The methods have been tested previously on the small set of reactions known as BH6,\cite{lynch2003small} where it was found that both PZSIC and LSIC improve the calculated reaction barriers over uncorrected DFAs.\cite{doi:10.1063/1.5129533} Here we extend the tests to the larger BH76 benchmark set.  
Compiled  by Truhlar and coworkers, BH76 \cite{zhao2005multi,zheng2009dbh24} includes forward and reverse barriers for 19 hydrogen transfer (HT) and 19 non-hydrogen transfer (NHT) reactions. (The individual sets are known as HTBH38 and NHTBH38.)  These reactions are important in chemical processes such as fuel combustion\cite{li2016theoretical} and catalysis, as well as in biological processes such as protein denaturation\cite{hynes2007hydrogen} and the design of enzymes.\cite{planas2021computational} 
The BH76 set has been used previously\cite{zheng2009dbh24} to benchmark other methods and has been updated over the years and incorporated into more comprehensive benchmark sets.\cite{zhao2008density,goerigk2017look,verma2019revised}  Below we report results using PZSIC and two LSIC methods that use different schemes for scaling the SIC.\cite{romero2020local,doi:10.1063/5.0004738}   

In recent years, the analysis of DFA errors has focused on separating the roles played by the approximate nature of the self-consistent DFA density and the limitations of the energy functional itself.\cite{Janesko2008,VERMA201210,Burke2013,Burke2015,Wasserman2017}
A density-driven error is indicated when a DFA prediction is significantly improved by evaluating  the density functional energy using the exact density of the many-electron system, rather than the self-consistent DFA density. 
In assessing density-driven errors, the Hartree-Fock (HF) density is often used in place of the exact density,\cite{sim2018quantifying} due to the high computational cost of obtaining the latter through high-level quantum chemistry calculations.  The HF density is like the exact density in that both stem from methods that are free from effects of electron self-interaction. Using HF densities in DFT calculations has proven remarkably successful  for predicting physical properties  in some contexts.\cite{paesani-water} 

Density driven errors tend to be especially large in systems with stretched bonds.\cite{Wasserman2017} Janesko and Scuseria found that semilocal DFA predictions for the barrier heights of the BH76 set are improved by using HF densities to evaluate the DFA energies (DFA@HF), as well as by using the self-consistent densities from hybrid functionals.\cite{Janesko2008}  By contrast, the corresponding DFA reaction energies are essentially unchanged in DFA@HF calculations.\cite{Janesko2008}  This indicates that it is the transition states that are sensitive to density, as had been recognized earlier.\cite{patchkovskii2002improving}  
Similarly, Verma, Perera, and Bartlett used DFA@HF to calculate both transition state geometries and barrier heights and they also found a significant improvement over the predictions of the self-consistent DFAs.\cite{VERMA201210}
Because the self-consistent application of PZSIC results in a self-interaction-free electron density, it is natural to ask whether evaluating DFA energies using PZSIC densities (DFA@PZSIC) also improves the DFA barrier heights. We show below that, in most cases, LDA@PZSIC indeed produces better reaction barriers, but the correction due to the inclusion of SIC is a larger effect and results in greater improvement. 

\section{Methodology}
We apply PZSIC and LSIC within the Fermi-L\"owdin orbital self-interaction correction (FLOSIC) scheme.\cite{doi:10.1063/1.4869581} In FLOSIC,\cite{doi:10.1063/1.4869581} the SIC correction to the total energy in Eq.~(\ref{eq:pzsic}) is computed using  local orbitals based on Fermi orbitals (FOs).\cite{Luken1982} The Fermi orbitals are obtained from the spin density matrix and spin density as
\begin{equation}
    F_{j\sigma}(\vec{r}) = \frac{\sum_i \psi_{i\sigma}(\vec{a}_{j\sigma}) \psi_{i\sigma}(\vec{r})}{\sqrt{\rho_{\sigma} (\vec{a}_{j\sigma})}} 
\end{equation}
where $i$ and $j$ are orbital indices, $\sigma$ is the spin index, and $\vec{a_i}$ are special points in space referred as Fermi orbital descriptor (FOD) positions. The normalized, but not mutually orthogonal, FOs are orthogonalized using the L\"owdin symmetric orthogonalization method,\cite{lowdin1950non} resulting in the orthornormal Fermi-L\"owdin orbitals (FLOs). The SIC methods used in this work require computation of orbitalwise  Coulomb and exchange potentials and thereby significantly raises the computational cost over DFAs. However, computation of each orbital dependent potential is completely independent, allowing easy  and efficient parallelization  over the orbitals.
Further details regarding the FLOSIC methodology and examples of FLOSIC calculations for various properties are available in Refs.~\citenum{doi:10.1063/1.4869581,doi:10.1063/1.4907592,PEDERSON2015153,PhysRevA.95.052505,doi:10.1063/1.4947042,doi:10.1002/jcc.25586,doi:10.1063/1.5120532,PhysRevA.103.042811, diaz2021implementation,carlos_dcep,aquino2020fractional}. 

We have used two versions of LSIC in this work that differ in the details of how the SIC is scaled. In LSIC($z$), the total energy is given as
 \begin{equation}
    E^{LSIC}[\rho_\uparrow,\rho_\downarrow] = E^{DFA}[\rho_\uparrow,\rho_\downarrow] - \sum_{i\sigma}^{occ}\{ U^{LSIC}[\rho_{i\sigma}] + E^{LSIC}_{XC}[\rho_{i\sigma},0] \}
\end{equation}
where
\begin{equation}
    U^{LSIC} [\rho_{i\sigma}] = \frac{1}{2}\int d^3r\, \, z_\sigma(\vec{r})
    \rho_{i\sigma(\vec{r})} \, \int d\vec{r'} 
    \frac{\rho_{i\sigma}(\vec{r^\prime})}{|\vec{r}-\vec{r^\prime}|},\\
\end{equation}
    
\begin{equation}
     E_{XC}^{LSIC} [\rho_{i\sigma},0] = \int d^3{r}\, \, z_\sigma(\vec{r})
     \rho_{i\sigma(\vec{r})}  \,  \epsilon_{XC}^{DFA} ([\rho_{i\sigma},0],\vec{r})
\end{equation}
where $\epsilon_{XC}^{DFA}$ is the exchange-correlation energy density. Here the scaling factor is the iso-orbital indicator $z_{\sigma}(\vec{r})=\tau_\sigma^W(\vec{r})/\tau_\sigma(\vec{r})$, where $\tau_\sigma^W = |\nabla \rho_{\sigma}|^2/8 \rho_{\sigma}$ is the von Weizacker kinetic energy density and $\tau_\sigma(\vec{r}) = \Sigma_i |\nabla \psi_{i,\sigma}|^2$,  lies between 0 and 1 and indicates the nature of the charge density at $\vec{r}$. $z_\sigma(\vec{r})=1$ for a density corresponding to a single electron orbital and $z_\sigma(\vec{r})=0$ where the density is uniform. Scaling the self-interaction correction terms with $z_\sigma$ thus retains the full correction for a one-electron density, making the theory exact in that limit, and eliminates the correction in the limit of uniform density where $E_{XC}^{DFA}$ is already exact by design.   

LSIC($w$) is a variation\cite{romero2020local} of the LSIC method where the scaling factor $z_{\sigma}(\vec{r})$ is replaced by an orbital dependent factor $w_{i \sigma}(\vec{r}) =\rho_{i\sigma} (\vec r)/\rho_{\sigma} (\vec{r}) $. Like $z_{\sigma}$, $w_{i\sigma}$ also goes to one in the one-electron limit.  In the slowly-varying, many-electron limit, $w_{i\sigma}$ does not vanish, but is expected to be small. Unlike $z_{\sigma}$, $w_{i\sigma}$ does not vanish near critical points of the charge density, for example at bond centers. Earlier tests done with LSIC($w$) showed that it is an improvement over PZSIC in the description of many properties.\cite{romero2020local} It is notably successful in predicting the binding energies of water clusters, where LSIC($z$) performs poorly. 

The LSIC reaction barriers and reaction energies reported here were evaluated perturbatively, using the self-consistent PZSIC densities.  We checked that using self-consistent LSIC densities to evaluate barriers and reaction energies for a small subset of the reactions gives essentially the same values as the perturbative approach.

\section{\label{sec:compdetails} Computational Details}
All of our calculations were performed using the FLOSIC code, \cite{FLOSICcode} which is based on the NRLMOL code.\cite{pederson1990variational,jackson1990accurate} The calculations were done at the all-electron level, using an extensive Gaussian basis set optimized for the PBE functional.\cite{PhysRevA.60.2840} For  first row atoms, a typical basis includes 5 s, 4 p, and 3 (Cartesian-type) d functions (543), based on between 12 and 15 single Gaussian orbitals (SGOs).  For a second row atom, a typical basis is (653) based on 16 to 18 SGOs.

We used the LDA as parameterized in the PW92 functional \cite{PhysRevB.45.13244} for all the calculations reported here. We computed LSIC-LDA total energies using the self-consistent FLOSIC-LDA density and optimized FODs. A self-consistency tolerance for the total energy of $10^{-6}$ $E_h$ was used in all calculations. The FOD positions were optimized until the forces on the FODs dropped below $10^{-3}$ $E_h$/$a_0$. 

We used the geometries given in the GMTKN55 database\cite{goerigk2017look} for the reactants (R), products (P), and transition states (TS) of the individual reactions of the BH76 set. These geometries  were optimized by Truhlar and coworkers at the quadratic configuration interaction with single and double excitations (QCISD) level of theory, using the modified G3 large basis set (MG3).\cite{zhao2005multi,zheng2009dbh24} The barrier heights for the forward (f) and reverse (r) reactions were computed from the total energies of the reactants (R), transition states (TS), and products (P) as 
\begin{eqnarray}
    \Delta E_\mathrm{f} = & E_\mathrm{TS} - E_\mathrm{R}\\
    \Delta E_\mathrm{r} = & E_\mathrm{TS} - E_\mathrm{P}
\end{eqnarray}
 
Reference values for the barrier heights were also taken from the GMTKN55 database.\cite{goerigk2017look} Consistent with these references, our calculations do not include zero-point energy corrections and contributions from spin-orbit interactions are neglected. 

We also calculated the barrier heights with LDA@PZSIC energies. This can be done conveniently by simply using the $E^{DFA}$ component of Eq.~(1) from a self-consistent PZ-SIC calculation.  We can then separately assess how the DFA reaction barriers change due to changes in the charge density (from the LDA@PZSIC terms) and compare that to the total change due to using {$E^{PZSIC}$}.

\section{Results and Discussions}
\subsection{Hydrogen-transfer reaction barriers}
\begin{figure}[!]
\includegraphics[width=1\columnwidth]{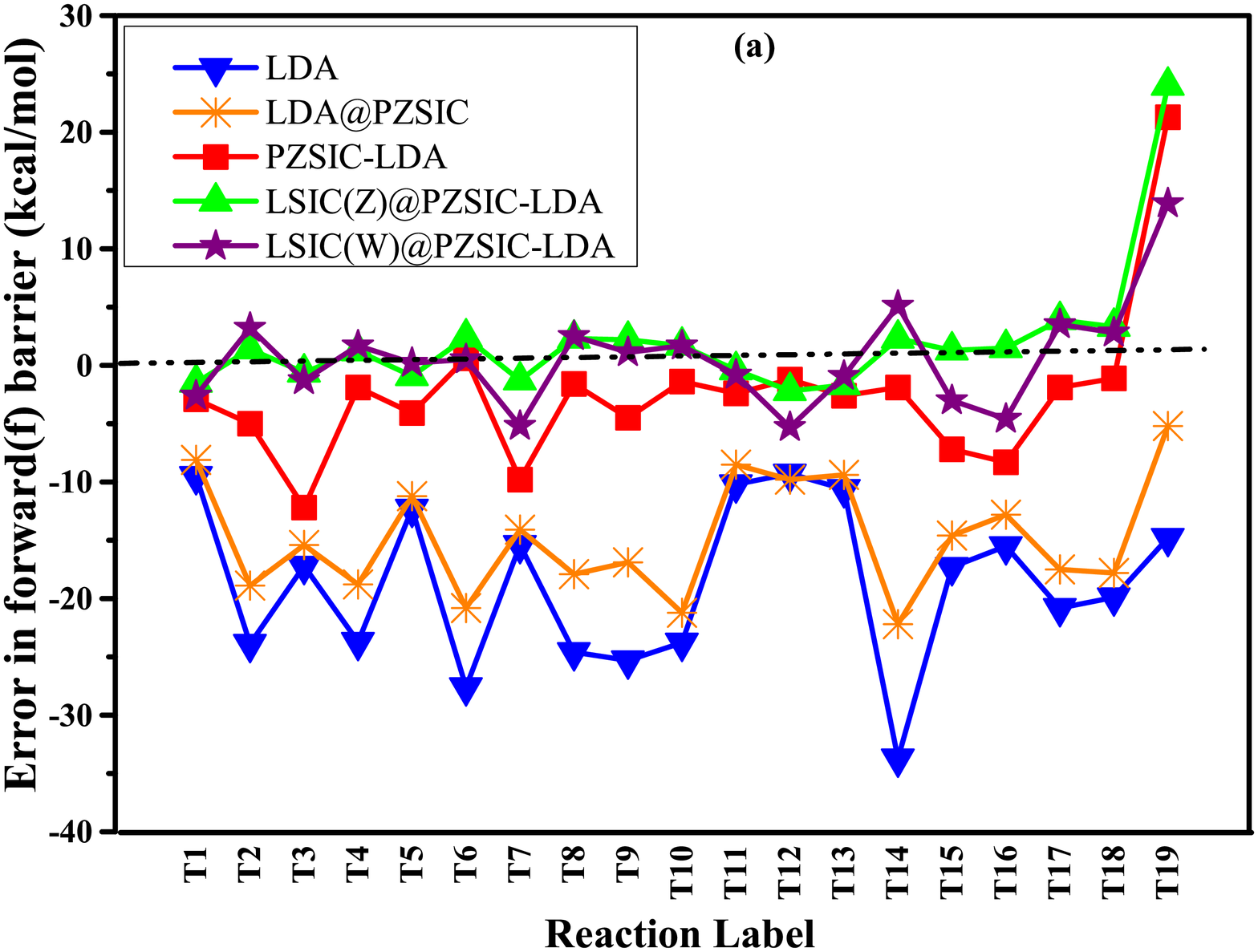}\\
\includegraphics[width=1\columnwidth]{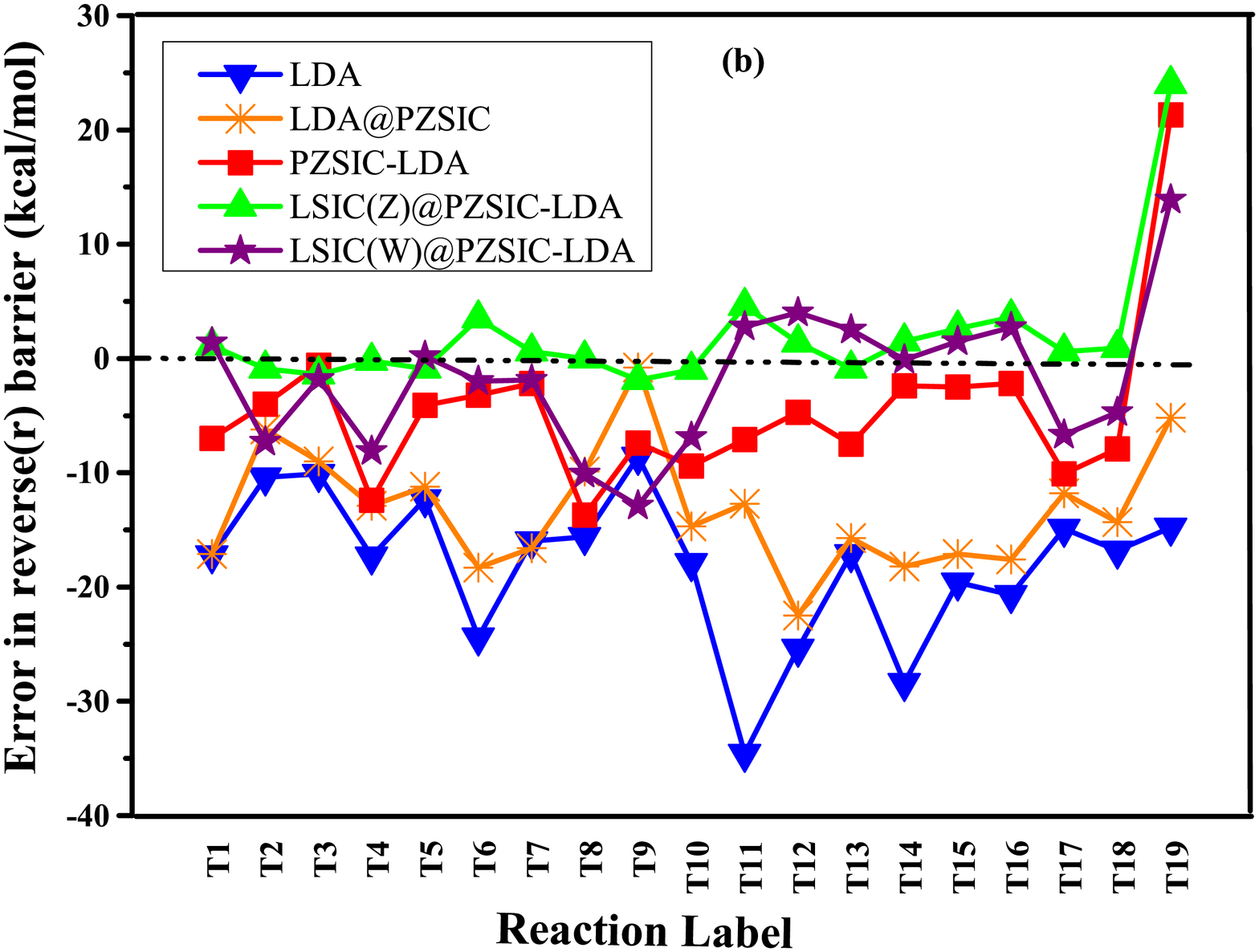}
\caption{\label{fig:HTBH} Errors in kcal/mol for (a) forward and (b) reverse barriers with respect to the reference values for hydrogen transfer reactions.}
\end{figure}

The values of the forward (f) and reverse (r) barriers calculated with LDA, PZSIC with LDA (PZSIC), LDA@PZSIC, LSIC($z$), and LSIC($w$) for the HTBH38 set are presented in Table \ref{tab:HTBH}, along with reference values.\cite{goerigk2017look} We also included the B3LYP values from Ref. \onlinecite{doi:10.1021/ct900489g} for comparison. Since the LDA functional was used for all SIC methods, we will not refer to the functional when discussing the SIC methods below. Errors relative to the reference values ($\Delta E_{\text{comp}} - \Delta E_\text{{ref}}$) for the forward and reverse barriers of each reaction are presented graphically in Fig. \ref{fig:HTBH}.  Mean errors (ME) and mean absolute errors (MAE) are summarized in Table \ref{tab:HTBH}.

\begin{longtable*}{@{\extracolsep{\fill}}cccrrrrrrr}
\caption{\label{tab:HTBH} The calculated forward and reverse reaction barriers heights (in kcal/mol) for the set of hydrogen-transfer reactions using the methods discussed in the text.}\\
\toprule
Label &Reaction & Barrier & LDA&LDA@PZSIC&PZSIC&LSIC($z$)&LSIC($w$)&B3LYP$^\text{a}$&Reference$^\text{b}$\\  
\midrule
T1 &H + HCl$\,\to\,$H$_2$ + Cl &f & -3.4&-2.0&3.2&4.6&3.5 &-0.87  &6.1\\
&& r& -9.2&-9.1&1.0&9.1&9.4 &4.6&8.0\\
T2 &OH + H$_2 \,\to\,$H$_2$O + H & f &-18.7&-13.7&0.2&6.6&8.1&0.9		&5.2\\
&& r& 11.2&15.4&17.6&20.7&	14.3&13.0 	&21.6\\
T3 & CH$_3$ + H$_2 \,\to\,$CH$_4$ + H  & f &-5.3&-3.5&-0.3&11.3&10.6&9.0		&11.9\\
&& r& 4.9&6.0&14.4&13.6&13.2&9.4		&15.0\\
T4 & OH + CH$_4 \,\to\,$H$_2$O + CH$_3$& f &-17.4&-12.5&4.4&7.5&7.7&2.4		&6.3\\
&& r& 2.2&6.6&7.1&19.3&11.4&14.1		&19.5\\
T5 & H + H$_2 \,\to\,$H$_2$ + H  & f &-2.6&-1.5&5.6&8.8&9.9&4.2		&9.7\\
&& r& -2.6&	-1.5&5.6&8.8&9.9&4.2		&9.7\\
T6&OH + NH$_3\,\to\,$H$_2$O + NH$_2$ & f &-24.2&	-17.4&	4.0&	5.8&3.9&-2.2		&3.4\\
&& r&-10.7&-4.6&	10.5&	17.2&11.7&7.3		&13.7\\
T7&  HCl + CH$_3 \,\to\,$CH$_4$ + Cl &f &-13.6&-12.3&	-8.0&0.5&-3.4&-1.1		&1.8\\
&& r& -9.2&-9.8&	4.6	&7.4&4.9&4.8		&6.8\\
T8& OH + C$_2$H$_6$ $\,\to\, $H$_2$O + C$_2$H$_5$&f & -21.1&	-14.4&	1.9&	5.8&6.0	&-0.5	&3.5\\
&& r& 4.8&10.5&6.7	&20.4&10.3&15.7		&20.4\\
T9 &   F + H$_2 \,\to\,$HF + H	 & f& -23.7&	-15.3&	-2.9&	3.8&2.7&-5.3		&1.6\\
&& r & 25.2	&33.0&	26.4&	31.9&20.9&23.0		&33.8\\
T10 &  O + CH$_4 \,\to\,$OH + CH$_3$ &f&-11.2&-6.8&	13.0&	16.1&16.1&7.7		&14.4\\
 && r & -9.0&	-5.8&	-0.5&	7.9&2.0&4.6		&8.9\\
T11 & H + PH$_3 \,\to\,$H$_2$ + PH$_2$&f& -7.3&	-5.6&	0.5&	2.5&2.1&-1.1		&2.9\\
 && r &-9.9	&12.0	&17.6	&29.3&27.5&23.2		&24.7\\
T12& H + HO$\,\to\,$H$_2$ + O& f& 1.6&	1.1&	9.7&	8.7&5.6&27.1		&10.9\\
 && r &-14.1&-9.3&	8.5&	14.6&17.2&29.7		&13.2\\
T13& H + H$_2$S$\,\to\,$H$_2$ + HS  & f& -6.7	&-5.5&	1.3&	2.2&3.0&-0.6		&3.9\\
 && r &0.1&	1.5&	9.7&	16.3&19.7&16.1		&17.2\\
T14&O + HCl$\,\to\,$OH + Cl & f& -25.1&	-11.8&	8.5&	12.7&15.5&1.6		&10.4\\
 && r &-18.5&	-8.3&	7.5&	11.4&9.8&4.4		&9.9\\
T15 &  CH$_3$ + NH$_2 \,\to\,$CH$_4$ + NH & f& -8.4	&-5.7&	1.7&	10.2&5.9&6.3		&8.9\\
 && r &2.4&4.9	&19.5&	24.6&	23.5&17.4	&22.0\\
T16 &  C$_2$H$_5$ + NH$_2 \,\to\,$C$_2$H$_6$ + NH & f&-5.7&-3.0&	1.5&	11.3&5.2&8.4		&9.8\\
 && r &-1.3&1.8	&17.2&	23.0&22.1&14.9		&19.4\\
T17 &  NH2 + C$_2$H$_6 \,\to\,$NH$_3$ + C$_2$H$_5$ & f& -9.5&	-6.2&	9.4&	15.2&14.8&8.9		&11.3\\
 && r &2.9	&6.0&	7.7&	18.4&11.1&15.7	&17.8\\
T18 &   NH$_2$ + CH$_4 \,\to\,$NH$_3$ + CH$_3$ & f& -6.0	&-3.9&	12.8&	17.2&16.7&11.5		&13.9\\
 && r &0.1	&2.6	&9.0	&17.8&12.2&13.6		&16.9\\
T19 & s-trans cis-C$_5$H$_8\,\to\, $s-trans cis-C$_5$H$_8$  & f &24.9&	34.5&	61.0&	63.7&53.6&38.8		&39.7\\
&& r& 24.9&	34.5&	61.0&	63.7&53.6&38.8		&39.7\\
\midrule
ME & & &  -18.4 &   -14.0&-3.6   &2.0& -0.6 &-3.2&\\
MAE &&&   18.4 &    14.0 & 5.8  &2.8  &  4.0 &4.9\\
\bottomrule
\multicolumn{10}{l}{$^\text{a}$Reference \onlinecite{doi:10.1021/ct900489g}} \\
\multicolumn{10}{l}{$^\text{b}$References \onlinecite{goerigk2017look,verma2019revised}}
\end{longtable*}  

The ME in the LDA results is -18.4 kcal/mol and the MAE is 18.4 kcal/mol, showing that LDA  consistently underestimates the barrier heights. The MAE  is in good agreement with that cited by Janesko and Scuseria\cite{Janesko2008} (17.9 kcal/mol). This trend of under-estimation of barrier heights by LDA for reaction barriers involving HT was established in early applications.\cite{JOHNSON1994,patchkovskii2002improving} For the majority of the HT reactions the LDA barrier has an incorrect negative sign, indicating that the energy of the TS is spuriously lower than that of the R or P in LDA. PZSIC-LDA reverses this behavior, giving correct positive barriers for all reactions except T3, T7, T9, and T10, where one of the barriers remains negative. These reactions involve radicals and strongly electronegative atoms.  Overall, the application of PZSIC improves the barriers significantly, although all but one of the barriers remain smaller than the reference values. The MAE drops to 5.8 kcal/mol for PZSIC which is higher than that for B3LYP (4.9 kcal/mol).

As seen in Table I and Fig.~1, the LDA@PZSIC barriers are also underestimated, although in most cases they are improved compared to LDA. Use of the PZSIC density drops the MAE from 18.4 kcal/mol for LDA to 14.0 kcal/mol for LDA@PZSIC. While this reduction of 4.4 kcal/mol is significant, it is only about one third of the total reduction of 12.6 kcal/mol due to PZSIC.  Unlike in PZSIC, the LDA@PZSIC barriers are still negative for nearly all the reactions for which the LDA barriers are also negative.  

The signs of the LSIC($z$) barriers are in agreement with the reference values for all the HT reactions and the errors are generally smaller than for PZSIC  and B3LYP. The  MAE  for LSIC($z$) is  2.8 kcal/mol. The LSIC($z$) barriers do not show the consistent underestimation seen for LDA and PZSIC-LDA in Fig.~1.  
The LSIC($w$) energy functional gives barrier heights that are similar to those of LSIC($z$) in many cases, but as different as 10 kcal/mol in others. On average, its performance lies between that of PZSIC and  LSIC($z$), with an  MAE of 4.0 kcal/mol.

The largest errors in the PZSIC and LSIC barriers occurs for the T19 reaction, which is an intra-molecular symmetric cis-trans isomerization involving breakage and formation of double bonds.In this case, the products and reactants are the same and therefore, the forward and reverse reaction barriers are the same. The SIC methods predict  barriers for T19 that are too large by 21.3, 24.0, and 13.9  kcal/mol for PZSIC, LSIC($z$), and LSIC($w$), respectively. The best performance for this reaction is given by LDA@PZSIC-LDA, which results in an error of -5.2 kcal/mol. The LSIC($z$) and LSIC($w$) MAE without including the T19 reaction are 1.6 and 3.4 kcal/mol, respectively.
All the reactions in Fig. \ref{fig:HTBH} except T19 involve H-abstraction
by a radical, where only a single bond is broken, accompanied by the formation of another 
single bond. The T19 reaction is 
more complicated, as it involves the simultaneous 
breaking and formation of several chemical bonds in the transition state. One possible 
reason for the poor performance of PZSIC in this case may be related to
the poor description of noded orbital densities by local and semi-local functionals
as discussed in Ref. \onlinecite{shahi2019stretched}.
In the double bond regions the local orbitals have more nodes to preserve orthogonality. 
We note that both PZSIC and the LSIC methods perform poorly in this case,
suggesting that the problem is not due to over-correcting for self-interaction in regions 
of slowly-varying density.  Furthermore, comparison of the LDA@PZSIC-LDA  and PZSIC results 
show that the large error in PZSIC arises mainly from the functional and not the density.  
Our methods are implemented with real orbitals. Using complex orbitals may alleviate the
problem as complex local orbitals can be less noded.

\subsection{Non-hydrogen transfer reaction barriers}  
\begin{figure}
\includegraphics[width=1\linewidth]{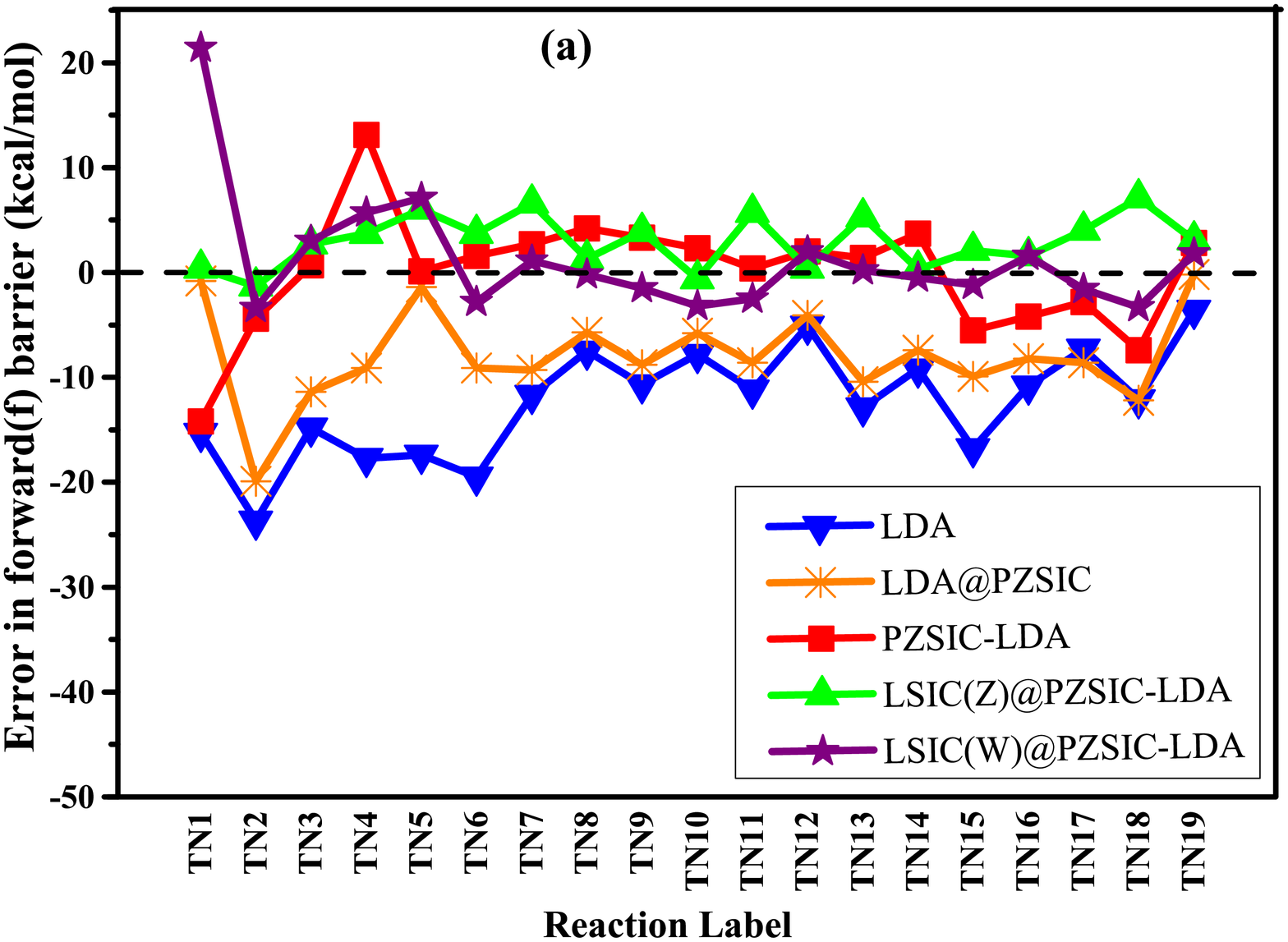}\\
\includegraphics[width=1\linewidth]{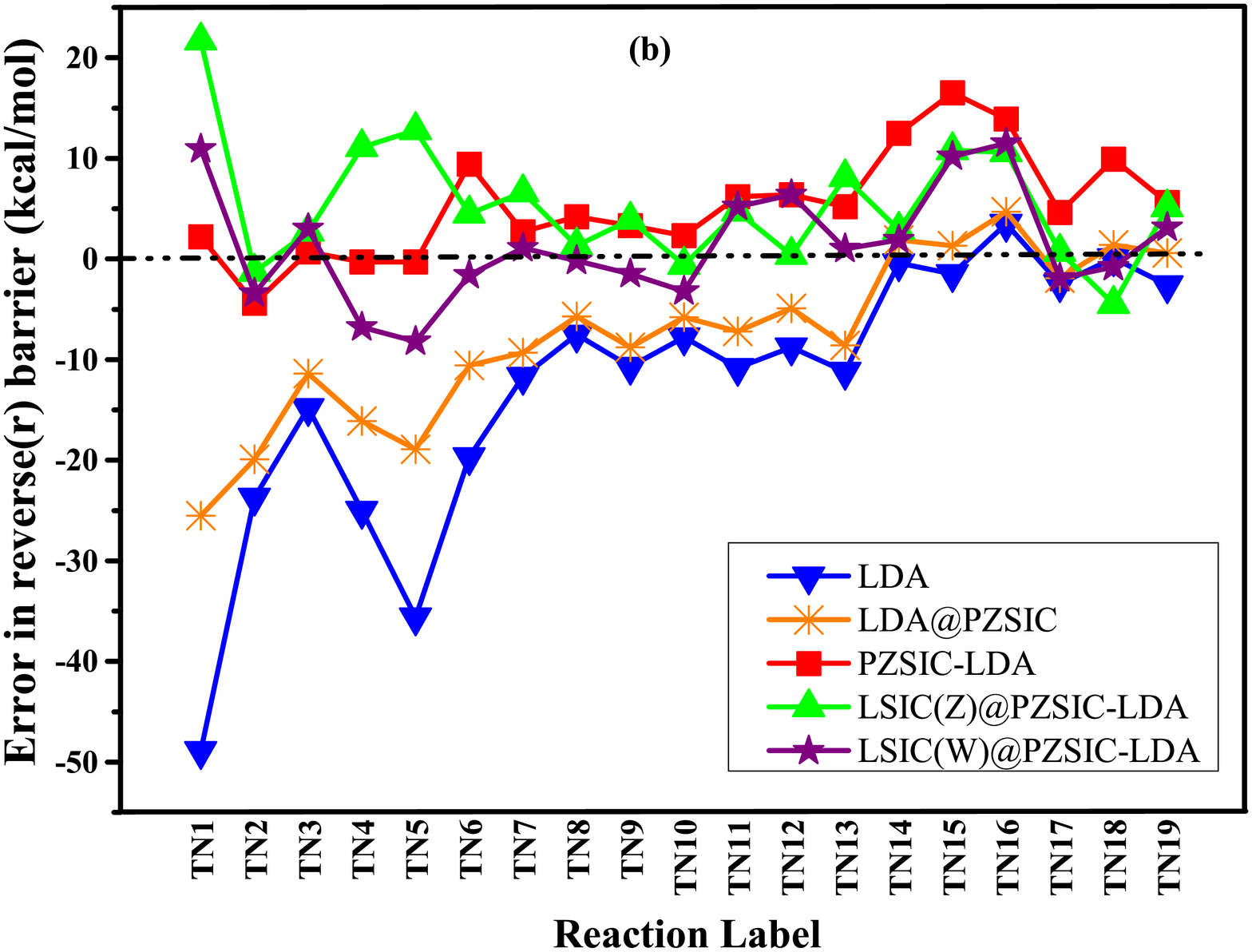}
\caption{\label{fig:NHTBH} Errors in kcal/mol for (a) forward and (b) reverse barriers with respect to the reference values for  non hydrogen transfer reactions.}
\end{figure}

The NHTBH38 set contains 6 heavy atom transfers, 8 nucleophilic substitutions, and 5 unimolecular and association reactions. These  are grouped together as TN1-TN6, TN7-TN14, TN15-TN19, respectively, in Table \ref{tab:NHTBH}, which compares the calculated barrier heights with reference values.\cite{goerigk2017look} It also shows the ME and MAE for the various methods.  The errors in the predicted barrier heights for each reaction are shown in Fig. \ref{fig:NHTBH}. 

\begin{longtable*}{@{\extracolsep{\fill}}cccrrrrrrr}
\caption{\label{tab:NHTBH} The calculated forward and reverse reaction barrier heights (in kcal/mol) for the set of non hydrogen-transfer reactions.}\\ 
\toprule
Label &Reaction & Barrier & LDA&LDA@PZSIC&PZSIC&LSIC($z$)&LSIC($w$)&B3LYP$^\text{a}$&Reference \\
\midrule
TN1 &H+N$_2$O$\,\to\,$OH+N$_2$ &f&2.4&16.9&3.5&18.1&39.1&11.4		&17.7\\
&&r& 33.7&	57.1&	84.8&	104.3&	93.5&73.1	&82.6\\
TN2 & H+FH$\,\to\,$H+FH & f &18.4&	22.2&	37.7&	40.8&	38.7&31.9	&42.1\\
&&r& 18.4	&22.2&	37.7&	40.8&	38.7&31.9	&42.1\\
TN3& H+ClH$  \,\to\,$HCl+H   & f &3.0&	6.4&	18.5&	20.5&20.8&12.4		&17.8\\
&&r& 3.0&	6.4&	18.5&	20.5&	20.8&12.4	&17.8\\
TN4 & H+FCH$_3 \,\to\,$HF + CH$_3$& f &12.8&	21.4&	43.6	&34.2&36.2&21.8		&30.5\\
&&r& 31.9&	40.8&	56.6&	68.0&	50.1&49.0	&56.9\\
TN5 & H+F$_2 \,\to\,$HF+F  & f &-15.9&	0.1&	1.6&	7.5&8.6	&-7.1	&1.5\\
&&r& 69.2&	85.9&	104.5&	117.6&96.6&95.2		&104.8\\
TN6&CH$_3$+FCl$\,\to\,$CH$_3$F+Cl & f &-12.4&	-2.0&	8.7&	10.8&4.3&-0.4		&7.1\\
&&r&40.1&	49.2&	69.2&	64.3&58.2&50.8		&59.8\\
TN7&  F$^-$+CH$_3$F$\,\to\,$FCH$_3$ + F$^-$&f &-12.3&	-9.9&	2.1&	6.0& 0.5	&-5.8	&-0.6\\
&&r& -12.3&	-9.9&	2.1&	6.0& 0.5&-5.8		&-0.6\\
TN8&  F$^-$$\cdots$CH$_3$F$\,\to\,$FCH$_3$$\cdots$F$^-$&f &5.9&	7.7&	17.6&	14.7&13.2&9.4		&13.4\\
&&r& 5.9&	7.7&	17.6&	14.7&13.2&9.4		&13.4\\
TN9 &  Cl$^-$+CH$_3$Cl$\,\to\,$ClCH$_3$+Cl$^-$ & f& -8.2&	-6.3&	5.8&	6.4&	1.0&-1.3	&2.5\\
&&r &-8.2	&-6.3&	5.8&	6.4&1.0	&-1.3	&2.5\\
TN10 & Cl$^-$$\cdots$CH$_3$Cl$ \,\to\,$ClCH$_3$$\cdots$Cl$^-$ &f&5.7&	7.7&	15.8&	12.9&10.3&8.8		&13.5\\
 &&r &  5.7&	7.7&	15.8&	12.9&	10.3&8.8	&13.5\\
TN11 & F$^-$+CH$_3$Cl$\,\to\,$FCH$_3$+Cl$^-$&f& -23.5&	-20.9&	-11.9&	-6.6&	-14.8&-18.4	&-12.3\\
 &&r &-9.0&	12.6&	26.0&	24.5&	25.0&18.1	&19.8\\
TN12& F$^-$$\cdots$CH$_3$Cl$\,\to\,$FCH$_3$$\cdots$Cl$^-$& f& -1.6&	-0.6&	5.5&	3.9&5.5	&-0.4	&3.50\\
 &&r &-1.6&	24.7&	36.0&	30.0&36.0&26.8		&29.6\\
TN13&OH$^-$+CH$_3$F$\,\to\,$HOCH$_3$+F$^-$  & f& -15.6&	-13.1&	-1.3&	2.6&-2.4&-8.6		&-2.70\\
 &&r &6.4	&9.0&	22.8&	25.7&18.6&12.5		&17.6\\
TN14&OH$^-$$\cdots$CH$_3$F$\,\to\,$HOCH$_3$$\cdots$F$^-$ & f& 1.8&	3.6&	14.7& 11.3&	10.4&6.0	&11.0\\
 &&r &47.3&	49.6&	60.2&	50.6&	49.6&45.6	&47.7\\
TN15 &   H+N$_2 \,\to\,$HN$_2$ & f& -2.2&	4.7&	9.1&	16.7&	13.5&7.4	&14.6\\
 &&r &9.4&	12.2&	27.4&	21.7&	21.1&10.7	&10.9\\
TN16 &  H+CO$\,\to\,$HCO & f&-7.6&	-5.0&	-1.0&	4.8&	4.8&-0.7	&3.2\\
 &&r &26.3&	27.5&	36.7&	33.4&	34.3&24.3	&22.8\\
TN17 &  H+C$_2$H$_4 \,\to\,$CH$_3$CH$_2$ & f& -5.3&	-6.6&	-0.8&	6.0&	0.4	&-0.3 &2.0\\
 &&r &39.5&	40.1&	46.6&	42.7&	40.1&41.7	&42.0\\
TN18 &   CH$_3$+C$_2$H$_4 \,\to\,$CH$_3$CH$_2$CH$_2$ & f& -5.7&	-5.8&	-1.0&	13.5&	3.1&6.2	&6.4\\
 &&r &33.1&	34.4&	42.9&	28.5&	32.2&29.3	&33.0\\
TN19 &HCN$\,\to\,$HNC   & f &44.5	&47.9&	50.9&	51.2&	49.9&47.7	&48.1\\
&&r& 30.4&	33.6&	38.6&	38.1&	36.1&33.5	&33.0\\
\midrule
ME &&& -12.5& -7.8& 3.1 & 3.7 & 1.3&-4.6&\\
MAE &&& 12.7& 8.3 & 5.0 & 4.6 & 3.9&4.9&\\
\bottomrule   
\multicolumn{10}{l}{$^\text{a}$Reference \onlinecite{doi:10.1021/ct900489g}} \\
\end{longtable*}  

The negative reference reaction barriers in Table II for reactions TN7, TN11, and TN13, are due to choosing the reactants as an isolated anion and molecule, which raises the energy of the reactants above the energy of the transition state. When the reference state for the reactants is chosen as the ion-molecule complex, as in reactions TN8, TN12, and TN14, the barrier becomes positive. This artifact is discussed by Zhao \textit{et al.} in Ref. \onlinecite{doi:10.1021/jp045141s}.

We observe the same general trends for the NHT reactions as we found for the HT set, namely that the MAEs follow the trend LDA $>$ LDA@PZSIC  $>$ PZSIC $>$ LSIC. The ME with LDA is -12.5 kcal/mol and the MAE is 12.7 kcal/mol. Although the LDA underestimates most of the barriers,  the LDA barrier is larger than the reference value for two reactions. 

To get a more detailed picture we calculated  MAEs for the different subsets of the NHT reactions (cf. Table \ref{tab:subsets}). For LDA, the largest errors occur for the heavy atom transfer reactions with an MAE of 23.0 kcal/mol and the smallest for the unimolecular and association reactions, with an MAE of 6.1 kcal/mol. For the unimolecular and association reactions the LDA forward barriers are too low and have incorrect signs, but the reverse barriers are in very good agreement with reference values. The overall effect is a relatively small MAE.  
 
The PZSIC MAE for the NHTBH38 set is 5.0 kcal/mol, but an examination of the subsets shows that PZSIC performs best for the nucleophilic substitutions with an MAE of 3.9 kcal/mol and worst for the unimolecular and association reactions with an MAE of 7.3 kcal/mol, which is surprisingly larger than that for LDA.   Although the MAE is higher with PZSIC, inspection of Table \ref{tab:NHTBH} shows that the corrections are in the right direction in all but one case. The largest errors for PZSIC for this group of reactions arise for the reverse barriers of TN15 and TN16 reactions, which 
involve breaking triple bonds. While LDA predicts the wrong sign for the barriers of 11 of the NHTBH38 reactions, PZSIC gives an incorrect sign for only three (Table \ref{tab:NHTBH}). 

\begin{table*}
\caption{\label{tab:subsets} Mean absolute errors (in kcal/mol) in barrier heights of different reaction types. The reference values are taken from Ref. \onlinecite{goerigk2017look} }
\begin{tabular*}{0.98\textwidth}{@{\extracolsep{\fill}}ccccccc}
\toprule
Method & &\multicolumn{3}{c}{NHTBH38 subsets}\\ \cmidrule(lr){3-5}
  &  & Heavy-Atom   &Nucleophilic   &Unimolecular and  &  &   \\ 
  & HTBH38 &  Transfer  & Substitution  & Association & NHTBH38 & BH76  \\ 
  \midrule
LDA               & 18.4  & 23.0 & 9.1 & 6.1 & 12.7 & 15.5 \\
LDA@PZSIC         & 14.0  & 13.2 & 7.0 & 4.9 &  8.3 & 11.2 \\
PZSIC-LDA         &  5.8  &  4.3 & 3.9 & 7.3 &  4.9 &  5.4 \\
LSIC($z$)@PZSIC   &  2.8  &  6.0 & 3.3 & 5.0 &  4.6 &  3.7 \\
LSIC($w$)@PZSIC   &  4.0  &  6.4 & 2.0 & 3.7 &  3.9 &  3.9 \\
B3LYP$^\text{a}$  &  4.5  &  8.2 & 4.2 & 2.0 &  4.9 &  4.7 \\
LDA@HF$^\text{b}$ & 10.2  & 13.0 & 1.3 & 3.5 &  5.5 &  7.9 \\
\bottomrule

\multicolumn{7}{l}{
$^{\text{a}}$ Reference \onlinecite{goerigk2017look}}\\

\multicolumn{7}{l}{
$^{\text{b}}$ Reference \onlinecite{Janesko2008}}
\end{tabular*}
\end{table*}
  
LDA@PZSIC reduces the MAE for the barriers to 8.3 kcal/mol for the NHTBH38 set overall.  For the heavy atom transfers, nucleophilic substitutions, and unimolecular and association reactions, the MAE are 13.2, 7.0, and 4.9 kcal/mol, respectively.

For LSIC($z$) and LSIC($w$), the  MAEs for the NHTBH38 set are 4.6 and 3.9 kcal/mol, respectively. The LSIC errors for the NHT set are larger than those for the HT set and the overall improvement over PZSIC is smaller.   For the LSIC approaches, the  errors for the substitution and association reactions are reduced but increase for the heavy atom transfer reactions compared to PZSIC.  The most notable case is the reverse barrier for TN1 (H+N$_2$O$\rightarrow$OH+N$_2$) reaction where both LSIC methods give large errors of 11 and 22 kcal/mol for (LSIC($w$) and LSIC($z$), respectively.  LSIC also corrects for the sign errors in PZSIC for the unimolecular and association reactions. 

The overall trend seen in Tables I and II is that PZSIC improves the reaction barriers over LDA and LSIC improves it further. 
In case of the HT reactions, the LSIC approaches produce smaller MAEs than the hybrid GGA functional B3LYP (MAE = 4.5 kcal/mol, \cite{goerigk2017look} see Table III), whereas the MAE for PZSIC is larger than in B3LYP.  For the NHT reactions, all the SIC methods produce results comparable to or better than B3LYP (MAE = 4.96 kcal/mol\cite{goerigk2017look}).

\subsection{Reaction Energy}
Computed reaction energies are compared with reference values\cite{goerigk2017look} in Table \ref{tab:REHT} and \ref{tab:RENHT} for the HT and NHT reactions, respectively. We removed from our analysis the six substitution reactions for which the reaction energies are identically zero because the R and P are the same. The MAE for the reaction energies with LDA are 6.8 kcal/mol for HT and 9.3 kcal/mol for NHT reactions.  For PZSIC, the MAE  decreases to 5.9 kcal/mol for the HT reactions, but increases to 9.9 kcal/mol for NHT.
\begin{table*}
\caption{\label{tab:REHT}Reaction energies  of the hydrogen-transfer reactions in kcal/mol for the different methods. The B3LYP and reference values are taken from Ref.  
 \onlinecite{goerigk2017look}.}
\begin{tabular*}{0.98\textwidth}{@{\extracolsep{\fill}}c c r r r r r r r}
\toprule
Label &Reaction &LDA&LDA@PZSIC&PZSIC&LSIC($z$)&LSIC($w$)&B3LYP&Reference \\
\midrule
T1 &H + HCl$\,\to\,$H$_2$ + Cl &5.8	&7.0	&2.2	&-4.5	&-5.9	&-5.5	&-1.9\\
T2 &OH + H$_2 \,\to\,$H$_2$O + H  &-29.9&	-29.1&	-17.4&	-14.1&	-6.0&	-12.0&	-16.4\\
T3& CH$_3$ + H$_2 \,\to\,$CH$_4$ + H  &-10.3&	-9.6&	-14.7&	-2.3&	-2.6&	-0.4	&-3.1\\
T4 & OH + CH$_4 \,\to\,$H$_2$O + CH$_3$ &-19.6&	-19.1&	-2.7&	-11.8&	-3.4&	-11.6&	-13.2\\
T6&OH + NH$_3\,\to\,$H$_2$O + NH2 &-13.5&	-12.7&	-6.5&	-11.4&	-7.8&	-9.5&	-10.3\\
T7&  HCl + CH$_3 \,\to\, $CH$_4$ + Cl  &-4.5&	-2.5&	-12.3&	-6.9&	-8.3&	-5.9	&-5.0\\
T8& OH + C$_2$H$_6$ $\,\to\, $H$_2$O + C$_2$H$_5$&-25.9&	-24.9&	-4.8&	-14.6&	-4.3&	-16.3	&-16.9\\
T9 &   F + H$_2 \,\to\,$HF + H	 & -48.9&	-48.2&	-29.3&	-28.1&	-18.2&	-28.4&	-32.2\\
T10 &  O + CH$_4 \,\to\,$OH + CH$_3$ &-0.4&	-0.9&	13.6&	8.2&	14.2&	3.1&	5.5\\
T11 & H + PH$_3 \,\to\,$H$_2$ + PH$_2$&-17.2&	-17.6&	-17.4&	-26.8&	-25.4&	-24.3	&-21.8\\
T12& H + HO$\,\to\,$H$_2$ + O&10.6&	10.5&	1.1&	-5.8&	-11.6&	-2.7&	-2.3\\
T13& H + H$_2$S$\,\to\,$H$_2$ + HS  &-6.9&	-7.0&	-8.4&	-14.1&	-17.0&	-16.6&	-13.3\\
T14&O + HCl$\,\to\,$OH + Cl &-4.8&	-3.5&	0.7	&1.3&	5.7&	-2.8&	0.5\\
T15 &  CH$_3$ + NH$_2 \,\to\,$CH$_4$ + NH &-10.8&	-10.6&	-17.9&	-14.5&	-17.6&-11.2&	-13.1\\
T16 &  C$_2$H$_5$ + NH$_2 \,\to\,$C$_2$H$_6$ + NH &-4.5&	-4.8&	-15.7&	-11.7&	-16.9&	-6.5	&-9.6\\
T17 &  NH2 + C$_2$H$_6 \,\to\,$NH$_3$ + C$_2$H$_5$ & -12.4&	-12.2&	1.7&	-3.3&	3.8&	-6.8&	-6.5\\
T18 &   NH$_2$ + CH$_4 \,\to\,$NH$_3$ + CH$_3$ & -6.1&	-6.4&	3.8&	-0.5&	4.5&-2.1	&-3.0\\
\midrule
MAE && 6.8&	6.7&	5.9&	2.3&	6.9&	2.2 &\\
\bottomrule          
\end{tabular*}
\end{table*}  
The HT reactions involve fragments with one electron densities where PZSIC works well. The corresponding B3LYP reaction energies\cite{goerigk2017look} are in better agreement with the reference values with MAEs of 2.2 and 2.7 kcal/mol for the HT and NHT sets, respectively. LSIC($z$) improves the reaction energies relative to PZSIC, with MAEs of 2.3 and 6.0 kcal/mol for the HT and NHT reactions, respectively.  Moreover, the signs 
of the LSIC($z$) reaction energies match those of the reference values except for TN15. 
These results suggest that some of the improvement to barrier heights with LSIC($z$) stems from the correction to the energies of equilibrium structures.
\begin{table*}
\caption{\label{tab:RENHT} Reaction energies in kcal/mol for the non hydrogen-transfer reactions with the  methods employed here are compared with reference values from Ref. \onlinecite{goerigk2017look}.}
\begin{tabular*}{0.98\textwidth}{@{\extracolsep{\fill}}c c r r r r r r r}
\toprule
Label &Reaction &LDA&LDA@PZSIC&PZSIC&LSIC($z$)&LSIC($w$)&B3LYP&Reference \\
\midrule
TN1 &H+N$_2$O$\,\to\,$OH+N$_2$ &-31.4&	-40.2&	-81.3&	-86.2&	-54.4&	-61.4 &-64.9\\
TN4 & H+FCH$_3 \,\to\,$HF + CH$_3$ &-19.1&	-19.4&	-13.0&	-33.8&	-14.0&	-26.9&-26.4\\
TN5 &  H+F$_2 \,\to\,$HF+F  &-85.1&	-85.9&	-102.9&	-110.1&	-87.9&	-101.7&-103.3\\
TN6&CH$_3$+FCl$\,\to\,$CH$_3$F+Cl &-52.5&	-51.1&	-60.4&	-53.5&	-53.8&	-51.3&-52.7\\
TN11 & F$^-$+CH$_3$Cl$\,\to\,$FCH$_3$+Cl$^-$&-32.5&	-33.6&	-37.9&	-31.0&	-39.9&	-36.1&-32.1\\
TN12& F$^-$$\cdots$CH$_3$Cl$\,\to\,$FCH$_3$$\cdots$Cl$^-$&-22.4&	-25.3&	-30.5&	-26.2&	-30.5&	-26.6&-26.1\\
TN13&OH$^-$+CH$_3$F$\,\to\,$HOCH$_3$+F$^-$  &-22.1&	-22.1&	-24.1&	-23.1&	-21.1&	-21.4&-20.3 \\
TN14&OH$^-$$\cdots$CH$_3$F$\,\to\,$HOCH$_3$$\cdots$F$^-$ &-45.5&	-46.0&	-45.5&	-39.3&	-39.1&	-40.1&-36.7\\
TN15 &   H+N$_2 \,\to\,$HN$_2$ &-11.5&	-7.5	&-18.3	&-5.0	&-7.7&	-3.5&3.7\\
TN16 &  H+CO$\,\to\,$HCO&-33.8&	-32.4&	-37.7&	-28.6&	-29.5&	-25.2&-19.6\\
TN17 &  H+C$_2$H$_4 \,\to\,$CH$_3$CH$_2$ & -44.8&	-46.7&	-47.4&	-36.7&	-39.7&	-42.0&-40.0\\
TN18 &  CH$_3$+C$_2$H$_4 \,\to\,$CH$_3$CH$_2$CH$_2$&-38.8&	-40.2&	-43.9&	-15.0&	-29.0&	-23.6&-26.6 \\
TN19 & HCN$\,\to\,$HNC  &14.1&	14.3&	12.3&	13.1&	13.8&	14.0&15.1\\
\midrule
MAE && 9.3&	8.4&	9.9&	6.0&	6.2&	2.7 & \\
\bottomrule         
\end{tabular*}
\end{table*}  
On the other hand, LSIC($w$) does not show a systematic improvement over PZSIC results. For HT reactions the LSIC($w$) MAE is 6.9 kcal/mol, much larger than for LSIC($z$) and comparable to that of PZSIC and LDA. But LSIC($w$) performs on par with LSIC($z$) and better than PZSIC and LDA for the NHT set (MAE = 6.2 kcal/mol). While our earlier work with LSIC($w$)\cite{romero2020local} showed improvement in predicting binding energies of water clusters compared to LSIC($z$), the results presented here show that using the iso-orbital indicator $z$ as the scaling factor is better suited for
predicting barrier heights and reaction energies. $z$ depends on the von Weiszacker kinetic energy density and vanishes in a region of uniform density. The scaling factor $w$, on the other hand, is the ratio of an orbital density to the total electron density.  It is expected to become small in uniform density regions, but does not necessarily vanish. Thus, $z$ can better identify uniform density regions and this may be the cause for the larger improvement of equilibrium properties using LSIC($z$) compared to LSIC($w$).

\subsection{LDA@PZSIC and density driven errors}
The results presented above allow us to assess the degree of density-driven error in the barrier heights and reaction energies predicted by the LDA@PZSIC results. MAEs for all of the methods discussed here are collected for convenience in Table III for the various groups of reactions in the BH76 set.  The MAEs for the hybrid B3LYP functional\cite{goerigk2017look} are shown for comparison, as are those for the LDA functional evaluated on HF densities (LDA@HF).\cite{Janesko2008}  For the HT reactions, use of LDA@PZSIC reduces the LDA MAE from 18.4 to 14.0 kcal/mol. Using LDA@HF reduces the LDA MAE to 10.2 kcal/mol.  Both LDA@PZSIC and LDA@HF indicate a significant density-driven contribution to the LDA error, but the LDA@HF result suggests a larger density-driven part.  Using PZSIC reduces the LDA error by a total of 12.6 to 5.8 kcal/mol. Taking into account the 4.4 kcal/mol reduction due to the density, 8.2 kcal/mol of the reduction could then be assigned to the PZSIC functional.  LSIC($z$) reduces the MAE a further 3.0 kcal/mol to 2.8 kcal/mol, implying an even larger improvement to the LDA energy functional.  For the NHT reactions, LDA@PZSIC reduces the LDA MAE by 4.4 kcal/mol to 8.3 kcal/mol, again indicating a sizable contribution of density-driven error.  LDA@HF reduces the LDA MAE by 7.2 kcal/mol to 5.5, again indicating a larger density-driven contribution than LDA@PZSIC.  For the NHT subsets, LDA@PZSIC and LDA@HF indicate essentially the same density-driven error for the heavy atom transfers, while LDA@HF suggests a somewhat larger density-driven contribution for the unimolecular reactions than LDA@PZSIC.  For the nucleophilic substitutions, LDA@HF suggests that nearly all the error is density-driven, while LDA@PZSIC predicts a much smaller density-driven contribution.  For the full BH76 set, LDA@PZSIC suggests a moderate density-driven error, reducing the LDA MAE from 15.5 to 11.2 kcal/mol.  LDA@HF reduces the LDA error to 7.9 kcal/mol, again indicating a larger density-driven contribution.
We point out that the full correction due to the use of the SIC energy functionals, whether PZSIC or LSIC($z$) or LSIC($w$), is clearly larger for the overall BH76 than the correction from either LDA@HF or LDA@PZSIC calculations. Comparison of the LDA@PZSIC results with LDA and PZSIC results shows that the functional driven errors are larger or comparable to the density driven errors depending on the type of reactions.

For the reaction energies, both LDA@PZSIC and LDA@HF indicate small density-driven errors.  For the HT reactions, the LDA MAE of 6.8 kcal/mol for the HT reactions is reduced to 6.7 kcal/mol  by LDA@PZSIC and 6.4 kcal/mol for LDA@HF.\cite{Janesko2008}  For the NHT reactions, the LDA and LDA@PZSIC MAEs are 9.3 and 8.4 kcal/mol.  The LDA and LDA@HF MAE's reported by Janesko and Scuseria are 6.7 and 4.6 kcal/mol, respectively.\cite{Janesko2008}
This value for the LDA MAE appears to differ from the present results due to a difference in how the averaging was done.  Including the six substitution reactions that have identically zero reaction energy in our averaging, we obtain an MAE of 6.4 kcal/mol for the LDA MAE, close to that reported by Janesko and Scuseria.\cite{Janesko2008}

\section{\label{sec:conclusion}Conclusions}
We examined the performance of SIC on reaction barrier heights and the reaction energies of the HT and NHT reactions in the BH76 benchmark set, using both the traditional Perdew-Zunger SIC (PZSIC) and locally scaled variations (LSIC($z$) and LSIC($w$)). The reaction barriers are strongly underestimated by LDA. The LDA errors in the barriers are significantly reduced by PZSIC and reduced further by the LSIC methods, especially for the HT reactions, where the MAE for LSIC($z$) is reduced to 2.8 kcal/mol (1.6 kcal/mol if one outlier reaction is removed from the averaging).  
We point out that the LSIC total energy values lie between LDA and LDA-PZSIC values.   Because the amount of energy corrections with LSIC are different for the reactant/product states and for the transition states,  LSIC barriers calculated from energy differences, do not necessarily result in values in between LDA and PZSIC. For the BH6 set of reactions, we observed that LSIC removed overcorrection of PZSIC more on the transition states than the reactant and product states such that the barrier heights become higher than the underestimated barriers of PZSIC.  
Overall, for the full BH76 set, LSIC($z$) produces an MAE of 3.7 kcal/mol compared to 4.7 kcal/mol for the popular B3LYP hybrid functional.\cite{goerigk2017look} 

Neither PZSIC nor LSIC($w$) improves the reaction energies of the BH76 set compared to LDA on average, but LSIC($z$) does, reducing the LDA MAE from 6.8 kcal/mol to 2.3 kcal/mol for the HT reactions and from 9.3 to 6.0 kcal/mol for the NHT set. We attribute the success of LSIC relative to PZSIC to the fact that it selectively scales down the SIC in regions where the density is slowly varying and the underlying LDA is expected to be worsened by the correction,\cite{santra2019perdew} but retains it at full strength in regions where the density is one-electron-like and SIC is needed most.\cite{doi:10.1063/1.5129533} The improvement is more consistent using the iso-orbital indicator $z$ for the scaling factor than for the orbital-based $w$ factor, because the former identifies slowly-varying regions better.

The self-interaction free density from a self-consistent PZSIC calculation can be used to probe for density-driven errors.  Results from LDA@PZSIC calculations are qualitatively similar to those of LDA@HF, with both approaches predicting that reaction barriers are subject to significant density-driven errors, but that the corresponding reaction energies are not. The full correction due to the use of the SIC energy functionals is, however, much larger for the overall BH76 set than the corrections from either LDA@HF or LDA@PZSIC calculations and shows significant corrections arising from the functional correction. The LDA@PZSIC results suggest smaller density-driven errors than LDA@HF on average.  

The LSIC method describes reaction barriers well when applied in conjunction with the LDA.  But the LDA is the simplest DFA and is relatively inaccurate for molecular properties.  It remains a challenge to develop an LSIC-like method for more accurate generalized gradient approximations (GGAs) and meta-GGAs.  The approach of Eq.~(3) encounters gauge inconsistencies for semi-local DFAs and repairing these leads to global rather than local scaling.\cite{bhattarai2020step} Finally, the largest errors registered here for the SIC methods involve breaking or forming multiple bonds.  Further study of these and related systems is needed to improve the performance of SIC methods.

\section*{Data Availability Statement}
The data that supports the findings of this study are available within the article.    

\begin{acknowledgments}
This work was supported by the US Department of Energy, Office of Science, Office of Basic Energy Sciences, as part of the Computational Chemical Sciences Program under Award No. DE-SC0018331. 
Support for computational time at the Texas Advanced Computing Center through NSF Grant No. TG-DMR090071, and at NERSC is gratefully acknowledged.
\end{acknowledgments}

\bibliography{bibtex} 

\end{document}